\RequirePackage[2020-02-02]{latexrelease}
\documentclass[12pt,titlepage]{article}
\usepackage{physics}
\usepackage{amsfonts}
\usepackage{amsmath}
\usepackage{amssymb}
\usepackage{amsthm}
\usepackage{setspace}
\usepackage{fullpage}
\usepackage{graphicx}
\usepackage{url}
\usepackage{sectsty}

\usepackage{hyperref}
\usepackage[numbers]{natbib}
\usepackage{indentfirst}
\usepackage{amsmath}
\usepackage{amsthm}
\usepackage{xcolor}
\usepackage{graphicx}
\usepackage{lipsum}
\usepackage{amssymb}
\usepackage{subfig}
\usepackage{cleveref}
\usepackage{amsmath}

\usepackage[utf8]{inputenc}
\usepackage{float}

\usepackage{color,colortbl}

\theoremstyle{plain}

\theoremstyle{definition}

\theoremstyle{remark}

\crefname{claim}{claim}{claims}
\Crefname{claim}{Claim}{Claims}
\crefname{app-corollary}{corollary}{corollaries}
\Crefname{app-corollary}{Corollary}{Corollaries}
\crefname{app-definition}{definition}{definitions}
\Crefname{app-definition}{Definition}{Definitions}
\crefname{figure}{figure}{figures}
\Crefname{figure}{Figure}{Figures}
\crefname{lemma}{lemma}{lemmata}
\Crefname{lemma}{Lemma}{Lemmata}
\crefname{app-lemma}{lemma}{lemmata}
\Crefname{app-lemma}{Lemma}{Lemmata}
\crefname{app-proposition}{proposition}{proposition}
\Crefname{app-proposition}{Proposition}{Proposition}
\crefname{app-theorem}{theorem}{theorems}
\Crefname{app-theorem}{Theorem}{Theorems}
\begin{document}

\setlength{\abovedisplayskip}{0pt}
\setlength{\belowdisplayskip}{0pt}
\setlength{\abovedisplayshortskip}{0pt}
\setlength{\belowdisplayshortskip}{0pt}



\title{A Comprehensive System for Secondary Structure Analysis of Protein Models}

\author{
Vedh Ramalingam Kannan
}
\vspace{0.5in}

\vspace{1in}


\def\submitdate{February 2024}

\date{\submitdate}

\begin{singlespace}
\maketitle
\end{singlespace}

\begin{abstract}
\begin{singlespace}

In protein structure analysis, the accurate characterization of secondary structure elements is crucial for understanding protein function and dynamics. This paper presents a software system designed for the comprehensive analysis of the secondary structure of protein models. Leveraging phi ($\phi$) and psi ($\psi$) torsion angles, the system utilises K-means clustering and outlier detection techniques to identify and classify folding structures within protein models. Through the visualisation of the Ramachandran plot, the software enables the differentiation of various secondary structure motifs, including alpha-helices, beta-sheets, and other structural elements. The incorporation of customisable threshold values facilitates the identification of outliers, providing insights into potential structural anomalies or misalignments within protein models. Overall, this software system offers researchers a powerful tool for comprehensive secondary structure analysis, to aid in enhancing understanding of protein structures created using both traditional methods such as X-Ray Diffraction and contemporary methods such as Artificial Intelligence.
\end{singlespace}
\vspace{1in}
\end{abstract}


\section{Background and Introduction} \label{sec:intro}

In protein structure analysis, the Ramachandran plot serves as a fundamental tool for evaluating the quality and validity of protein models by visually representing the distribution of phi ($\phi$) and psi ($\psi$) angles for amino acid residues in the protein backbone \cite{Ramachandran1963}. The $\phi$ angle signifies the rotation around the C-N bond, while the $\psi$ angle signifies the rotation around the C-C bond.

Different folding structures, including alpha-helices, beta-sheets, and other motifs, exhibit distinct patterns of torsion angles. For instance, the right-handed alpha-helix is characterised by specific ranges of $\phi$ and $\psi$ angles, approximately $-57°$ to $-47°$ and $-57°$ to $-47°$, respectively. Conversely, the left-handed alpha-helix displays similar ranges, albeit with positive values for both $\phi$ and $\psi$ angles. The $3_{10}$ helix and the pi helix also possess unique torsion angle preferences, with ranges of approximately $-74°$ to $-4°$ and $-57°$ to $-70°$, respectively.
In contrast, beta-sheets are characterised by different torsion angle distributions. The parallel beta-sheet exhibits $\phi$ and $\psi$ angles around -119° and +113°, respectively, while the anti-parallel beta sheet is characterised by ideal angles around $-139°$ and $+135°$ \cite{CHOUDHURI2014183}.

These regions, defined by the sterics of amino acid residues, are delineated by the Ramachandran plot and a large number of deviations of residues from these regions, commonly termed as Ramachandran outliers, may indicate potential issues with the protein model and necessitate further examination.

This study has developed research software aimed at simplifying the identification of various folding structures, offering a user-friendly approach to gaining in-depth insights into this secondary structure of proteins.

\section{Methods} \label{sec:desc}
\subsection{Data Preprocessing}

The raw data for this analysis is protein models in the *.pdb file format. A script in the software generates a *.tsv file of all  these files to extract $\phi$ and $\psi$ angles for each amino acid residue. The Bio.PDB module is then used to parse the PDB files and extract the required structural information, writing to file as shown below \cite{albert_bio}.

\begin{figure}[h]
    \centering
    \includegraphics[scale=0.4]{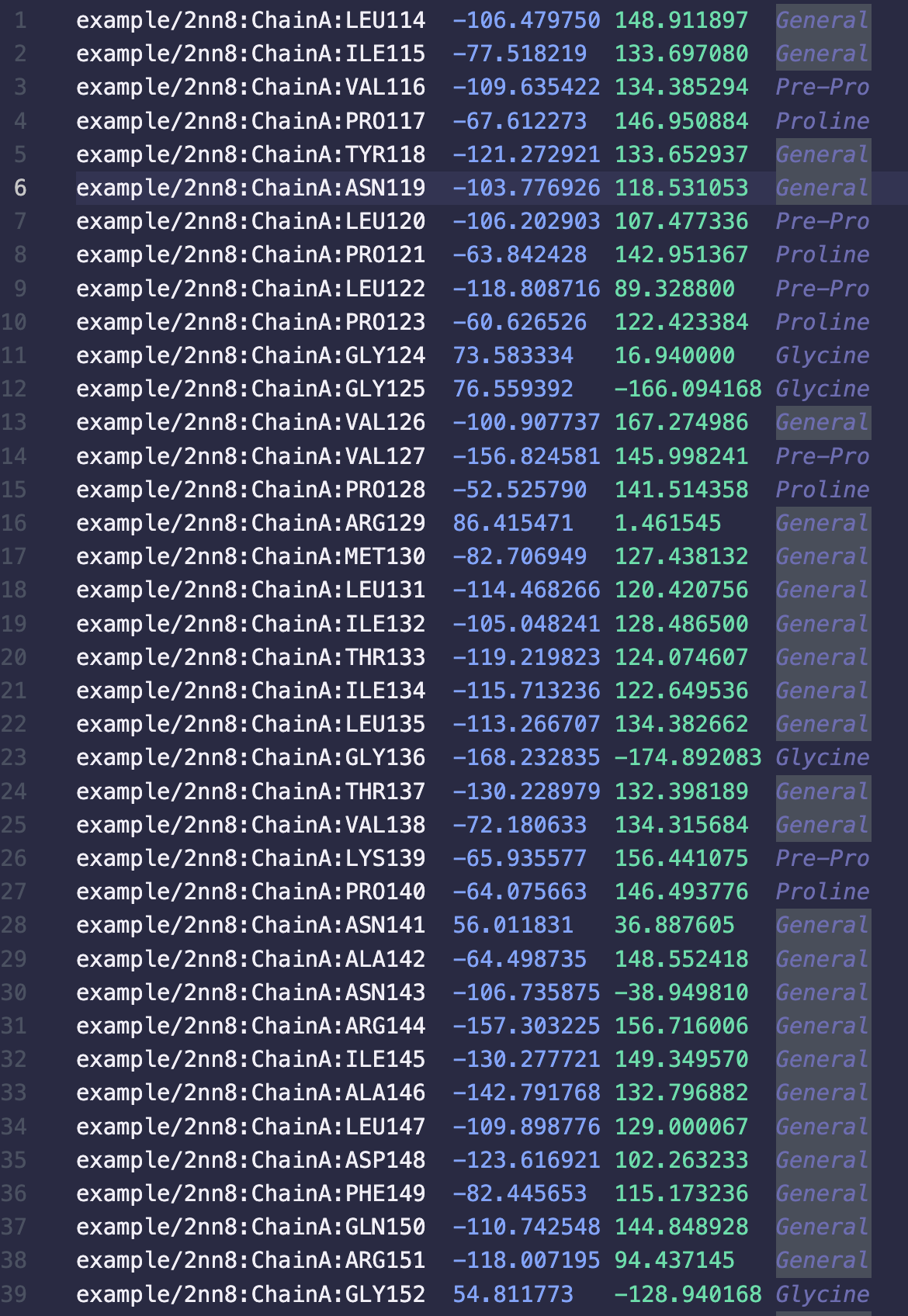}
    \caption{An example TSV file generated from a protein model in the RCSB databank}
    \label{fig:tsv-example}
\end{figure}

The above file shows each amino acid chain (column 1), the $\phi$ angles (column 2) and the $\psi$ angles (column 3) for each amino acid residue (column 4) in the protein model. All following computational analyses are performed on a file of this format.

\subsection{K-Means Clustering}
The core functionality of the software relies on the application of the K-means clustering algorithm to group $\phi$ and $\psi$ angles extracted from protein models into distinct clusters representing different folding structures. K-means clustering is a partitioning algorithm that aims to divide a dataset into K clusters, where each data point belongs to the cluster with the nearest mean, or centroid, and iteratively updates the centroids to minimise the within-cluster sum of squares. It is characterised by the formula:
\begin{equation}
\underset{\mathbf{C}}{\text{argmin}} \sum_{i=1}^{k} \sum_{\mathbf{x} \in S_i} ||\mathbf{x} - \mathbf{\mu}_i||^2
\end{equation}

\begin{itemize}
  \item \( k \) is the number of clusters.
  \item \( S_i \) is the set of data points assigned to cluster \( i \).
  \item \( \mathbf{\mu}_i \) is the centroid (mean) of cluster \( i \) 
  \item \( \mathbf{C} \) represents the set of cluster centroids.
\end{itemize}

\begin{figure}[H]
    \centering
    \includegraphics[scale=0.5]{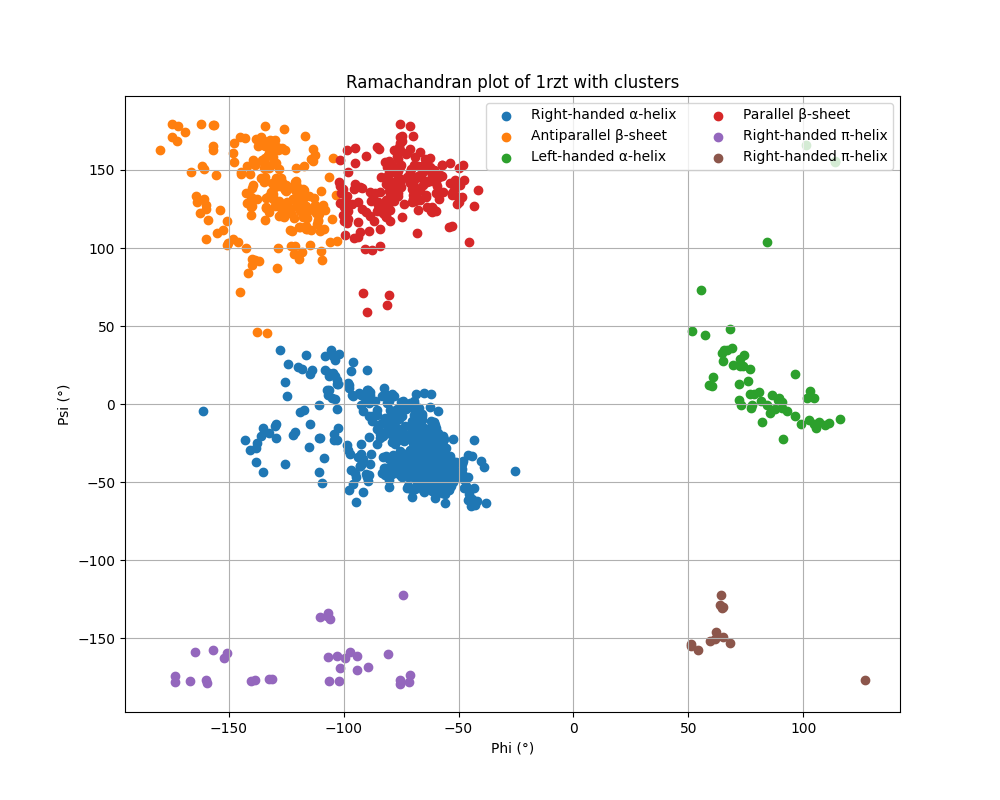}
    \caption{Ramachandran Plot of model 1rzt}
    \label{fig:RPlot_1rzt}
\end{figure}

In the context of protein structure analysis, the number of clusters (K) is determined based on the number of folding structures of interest. Each cluster represents a distinct folding structure characterised by specific ranges of $\phi$ and $\psi$ angles. For example, to identify right-handed alpha-helices, the software initialises K centroids corresponding to the ideal $\phi$ and $\psi$ angles of right-handed alpha-helices, as mentioned above \cref{sec:intro}. Similarly, centroids representing beta-sheets and other motifs are initialised accordingly. The clustering algorithm then iteratively assigns data points to the nearest centroids and updates the centroids to minimise the within-cluster sum of squares, effectively grouping the $\phi$ and $\psi$ angles into clusters corresponding to different folding structures. The outcome of this program is a plot as shown below in \cref{fig:RPlot_1rzt}

This plot was generated using model 1rzt from the Protein Data Bank \cite{Kunkel2004}. It is a model of the crystal structure of DNA polymerase lambda complexed with a two nucleotide gap DNA molecule.

\subsection{Outliers}
\begin{figure}[H]
    \centering
    \includegraphics[scale=0.5]{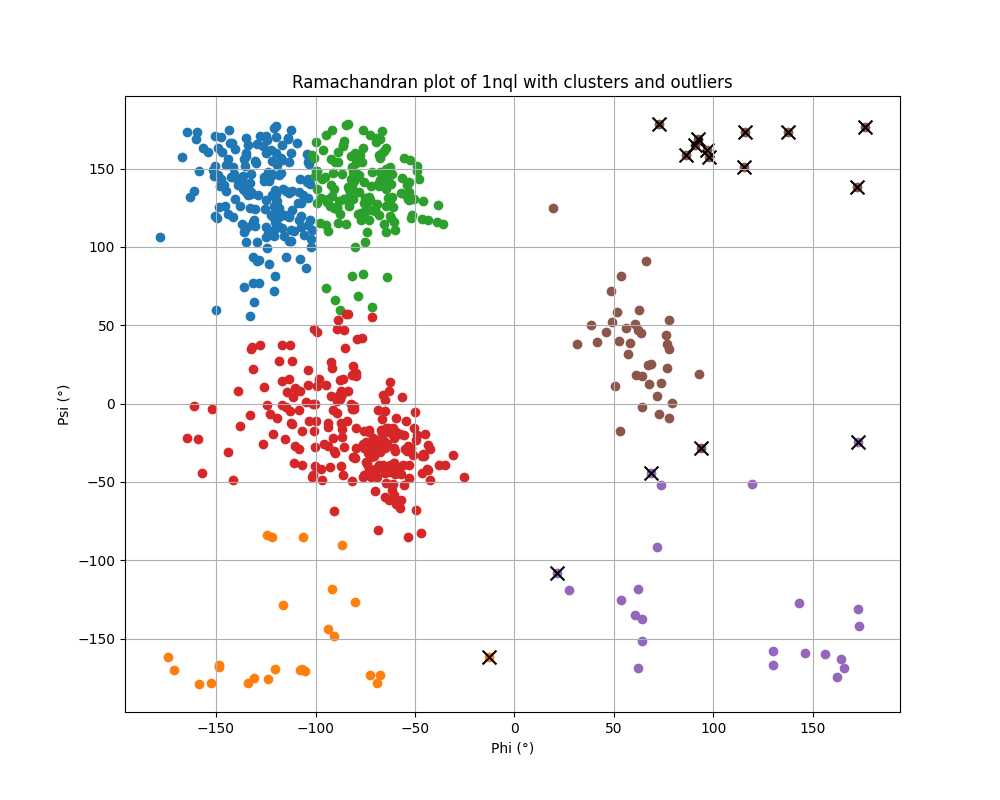}
    \caption{Ramachandran Plot of model 1nql}
    \label{fig:1nql}
\end{figure}

In addition to clustering, the software includes a feature for outlier detection to identify residues that deviate significantly from the expected torsion angle distributions. Outliers in the Ramachandran plot may indicate potential issues with the protein model, such as structural anomalies or misalignments. The software employs a distance-based approach to identify outliers, where residues that fall beyond a certain threshold euclidean distance from the cluster centroids are flagged as outliers with an 'x' mark. Notably, empirical observations have shown that setting the threshold within the range of $85 \leq x \leq 100$ yields results closely aligned with the validation reports provided by the Protein Data Bank (PDB) for the corresponding models. The outcome of introducing this into the program can be seen in \cref{fig:1nql}.

This plot was generated using model 1nql from the Protein Data Bank \cite{FERGUSON2003507}. It is a model of the structure of the extracellular domain of human epidermal growth factor (EGF) receptor in an inactive (low pH) complex with EGF.

\newpage

\section{Conclusion} \label{sec:conclusion}

This study has presented a research software aimed at enhancing the accuracy assessment of protein models through the analysis of phi ($\phi$) and psi ($\psi$) torsion angles. Through the utilisation of K-means clustering and outlier detection techniques, the software successfully identified and characterised folding structures within protein models, allowing for a more comprehensive understanding of their secondary structure.

By leveraging the Ramachandran plot, the software facilitated the visualization and differentiation of various folding phenomena, including alpha-helices and beta-sheets, providing researchers with valuable insights into protein conformation and stability. Additionally, the incorporation of outlier detection mechanisms enabled the identification of residues deviating significantly from expected torsion angle distributions, thereby pinpointing potential structural anomalies or misalignments within protein models.

\section{Data Availability}  \label{sec:fdesc}

All protein models used for testing in this study were either found on the Protein Data Bank (https://rcsb.org) or created using AlphaFold \cite{berman2000protein} \cite{jumper2021highly}. All software which was used can be found in the author's Github repository \cite{Kannan_Stereochemical-Assesments_2024}.

\begin{singlespace}
\bibliographystyle{style}

\bibliography{biblio}
\end{singlespace}

\end{document}